\documentclass[twocolumn,superscriptaddress,aps,pre]{revtex4-2}

\usepackage{graphicx}
\usepackage{hyperref}
\usepackage{amsmath,amssymb}
\usepackage[justification=raggedright,singlelinecheck=false]{caption} 
\usepackage{subcaption}
\usepackage{xcolor}
\begin{document}
	
	\title{Mittag-Leffler Quantum Statistics and Thermodynamic Anomalies }


\author {Maryam Seifi }
\email{m.seifi.j@uma.ac.ir}
\affiliation{Department of Physics, University of Mohaghegh Ardabili, P.O. Box 179, Ardabil, Iran}

\author {Zahra Ebadi}
\email{z.ebadi@uma.ac.ir}
\affiliation{Department of Physics, University of Mohaghegh Ardabili, P.O. Box 179, Ardabil, Iran}

\author {Hamzeh Agahi}
\affiliation{Department of Mathematics, Faculty of Basic Science, Babol Noshirvani University of Technology, Shariati Ave., Babol 47148-71167, Iran}

\author {Hossein Mehri-Dehnavi}
\affiliation{Department of Physics, Faculty of Basic Science, Babol Noshirvani University of Technology, Babol 47148-71167, Iran}

\author {Hosein Mohammadzadeh}
\email{mohammadzadeh@uma.ac.ir}
\affiliation{Department of Physics, University of Mohaghegh Ardabili, P.O. Box 179, Ardabil, Iran}

	\begin{abstract}
Building upon the framework established in our recent work [M. Seifi \textit{et al.}, \textit{Phys. Rev. E} \textbf{111}, 054114 (2025)], wherein a generalized Maxwell-Boltzmann distribution was formulated using the Mittag-Leffler function within the superstatistical formalism, we extend this approach to the quantum domain. Specifically, we introduce two statistical distributions—termed the Mittag-Leffler–Bose–Einstein (MLBE) and Mittag-Leffler–Fermi–Dirac (MLFD) distributions—constructed by generalizing the conventional Bose-Einstein and Fermi-Dirac distributions through the Mittag-Leffler function. This generalization incorporates a deformation parameter $\alpha$, which facilitates a continuous interpolation between bosonic and fermionic statistics, while inherently capturing non-equilibrium effects and generalized thermodynamic behavior. We analyze the thermodynamic geometry associated with these distributions and identify significant departures from standard statistical models. Notably, the MLBE distribution manifests a Bose-Einstein-like condensation even in the absence of interactions, whereas the MLFD distribution exhibits unconventional features, such as negative heat capacity in the low-temperature regime. These findings highlight the pivotal role of statistical deformation in determining emergent macroscopic thermodynamic phenomena.

\end{abstract}
\maketitle

	\section{Introduction}\label{1}
The Maxwell–Boltzmann (MB), Bose–Einstein (BE), and Fermi–Dirac (FD) standard distributions constitute the fundamental theoretical framework of equilibrium statistical mechanics. These distributions provide quantitatively accurate descriptions of the statistical behavior of particles in thermal equilibrium. Central to their formulation is the Boltzmann factor, $e^{-\beta \epsilon}$, where $\beta = 1/(k_B T)$ denotes the inverse temperature ( $k_B$ is the Boltzmann constant) and $\epsilon$ represents the energy of a microstate. This factor assigns the relative statistical weight of each microstate based on its energy. Although these standard distributions are highly effective under equilibrium conditions and for idealized systems, they often fail to capture the full complexity of systems exhibiting non-equilibrium behavior, strong correlations, or non-ideal interactions. One systematic approach to address these limitations involves generalizing the Boltzmann factor itself. Such generalizations yield extended statistical frameworks that are better equipped to describe a broader range of physical phenomena.

Several generalized statistical frameworks have been developed along these lines. Tsallis statistics replace the exponential with a power-law function derived from a nonextensive entropy, introducing a parameter $q$ that captures deviations from standard thermodynamics~\cite{jiulin2007nonextensivity,cohen2002statistics,plastino2004Tsallis,tamarit2005relaxation,Tsallis2002mixing}. Kaniadakis statistics introduce a deformation parameter $\kappa$ to construct a relativistically motivated generalization of the exponential function~\cite{kaniadakis2021new}. Superstatistics takes a different approach by averaging fluctuations in intensive parameters (such as temperature), thereby modeling systems as superpositions of local equilibria~\cite{beck2003superstatistics}. Each of these frameworks distinctly extends the traditional Boltzmann factor, allowing statistical mechanics to address more complex or non-equilibrium systems.

Motivated by efforts to generalize the Boltzmann factor, we employ the Mittag-Leffler (ML) function, a natural extension of the exponential function widely used to model complex systems exhibiting anomalous dynamics ~\cite{gorenflo2020mittag,dos2020mittag,mathai2010mittag}. The ML function features a tunable parameter, $\alpha$, which enables it to capture deviations from standard behavior and extend traditional statistical distributions. Building on this, in our earlier study ~\cite{seifi2025intrinsic}, we proposed the MLMB distribution, derived by generalizing the classical MB distribution through the expansion of the standard exponential function into the ML function. This generalization, parameterized by $\alpha$, enables a more comprehensive representation of particle interactions, encompassing behaviors not captured by the standard MB distribution. Notably, this formulation provides a continuous transition between classical and quantum statistical regimes, reproducing fermion-like and boson-like characteristics, thus effectively modeling interaction-induced deviations within a unified theoretical framework.

Building upon the ML generalization of the Maxwell-Boltzmann distribution \cite{seifi2025intrinsic}, the present study extends this framework to the quantum domain by introducing the Mittag-Leffler Bose-Einstein (MLBE) and Mittag-Leffler Fermi-Dirac (MLFD) distributions -ML-based distributions-. These novel distributions are formulated by generalizing the exponential function in the canonical BE and FD distributions to the ML function, thereby introducing the tunable parameter $\alpha$ into the quantum statistical weights. This extension preserves the essential quantum characteristics of bosons and fermions while allowing for a more flexible description of particle interactions and thermodynamic behavior, particularly in systems that deviate from idealized equilibrium conditions.

To investigate the physical implications of these generalized distributions, we employ the framework of thermodynamic geometry, which provides a powerful geometric interpretation of thermodynamic state space. By analyzing the thermodynamic curvature and associated metrics, this approach reveals information about microscopic interactions, stability, and phase transitions within the system. Thermodynamic geometry has previously provided important information on systems governed by generalized statistics such as Tsallis, Kaniadakis, and superstatistics~\cite{sagi2012observation,saporta2019self,colbrook2017scaling,pessoa2021information,mirza2009nonperturbative,mirza2010thermodynamic,oshima1999riemann}, and its application to the MLBE and MLFD distributions provides detailed information on their thermodynamic behavior, including heat capacity and condensation phenomena.

In the following sections, we develop these ideas in detail. 
Section~\ref{II} provides the broader context of generalized statistics and highlights the role of ML distributions in this framework. Section~\ref{3} introduces the ML-based distributions, specifically the MLBE and MLFD distributions. Section~\ref{4} discusses the thermodynamic quantities associated with these distributions—including internal energy, particle number—highlighting deviations from conventional statistics. Section~\ref{5} explores the geometric structure of the MLBE and MLFD statistics through thermodynamic metrics and curvature. Section~\ref{6} is devoted to the analysis of heat capacity, with particular emphasis on the emergence of negative heat capacity in the MLFD statistic and the evidence for such behavior in finite systems. Section~\ref{7} presents a comparative study of ML-based statistics with Tsallis and Kaniadakis distributions, highlighting the similarities and differences in their thermodynamic behavior. Section~\ref{8} applies MLBE statistics to the Debye solid, examining modifications to the heat capacity across both low- and high-temperature regimes, and providing a comparison with the Debye $q$-deformed model. Finally, Section ~\ref{9} summarizes the main findings, discusses potential applications, and outlines future research directions.

\section{Broader Context of Generalized Statistics and the Role of Mittag-Leffler Distributions}\label{II}

Many complex systems deviate strongly from the assumptions underlying the classical Boltzmann factor $e^{-\beta \epsilon}$, which presumes Markovian dynamics and exponential relaxation. In practice, relaxation is often non-exponential, correlations extend over long times, and transport exhibits anomalous scaling. In such contexts, the ML function arises naturally as the fundamental solution of fractional differential equations and provides the correct generalization of exponential weights. Its role in statistical mechanics is therefore not purely mathematical, but physically motivated by the ubiquity of fractional kinetics and memory effects.

For instance, glassy materials and disordered systems exhibit stretched-exponential relaxation of the form
\[
e^{-(t/\tau)^\beta},
\]
where $t$ denotes time, $\tau$ is a characteristic relaxation time, and $\beta$ ($0 < \beta < 1$) is the stretching exponent that quantifies deviations from Debye’s exponential law. At long times, such stretched exponentials asymptotically cross over into power-law decay \cite{bouchaud1992weak}. This crossover is precisely captured by the one-parameter ML function $E_\alpha(-t^\alpha)$, where $\alpha$ ($0<\alpha \leq 1$) is the fractional order parameter: for $\alpha = 1$ the function reduces to a pure exponential, while for $0<\alpha<1$ it exhibits algebraic tails.

Similarly, anomalous diffusion in porous media, biological cells, or turbulent plasmas is characterized by the mean-squared displacement
\[
\langle x^2(t) \rangle \sim t^\alpha,
\]
where $x$ is the displacement of a diffusing particle and the exponent $\alpha$ quantifies the deviation from normal Brownian diffusion (with $\alpha = 1$ corresponding to ordinary diffusion, $\alpha < 1$ to subdiffusion, and $\alpha > 1$ to superdiffusion). Such scaling laws emerge directly from fractional Fokker–Planck equations whose unique solutions are expressed in terms of ML functions \cite{metzler2000random, sandev2011fractional}.

Non-Markovian processes, including protein folding kinetics and earthquake aftershocks, can also be modeled by memory kernels. In the Laplace domain these kernels take the form
\[
\mathcal{L}\{t^{\alpha-1}E_{\alpha,\alpha}(-t^\alpha)\} = \frac{1}{s^\alpha + \lambda},
\]
where $\mathcal{L}$ denotes the Laplace transform, $s$ is the Laplace variable conjugate to time $t$, and $\lambda > 0$ is a characteristic rate parameter. This functional form arises naturally from fractional operators and cannot be represented by ordinary exponential kernels \cite{mainardi2020mittag}.

These examples illustrate that the ML function is far more than a mathematical curiosity: it is central to the physics of systems with long memory, anomalous transport, and hierarchical energy landscapes. Indeed, ML functions have long been recognized as the “queen function of fractional calculus,” emerging naturally as the fundamental solution of fractional-order differential equations that describe anomalous diffusion, non-exponential relaxation, and long-memory processes \cite{gorenflo2020mittag, haubold2011mittag}. Its relevance spans numerous areas of physics: modeling viscoelastic relaxation via the fractional Zener model, where stress $\sigma(t)$ and strain $\varepsilon(t)$ are related by fractional derivatives \cite{pritz2003five, nonnenmacher1991fractional}; describing anomalous transport and dispersion in porous or plasma environments via fractional advection–dispersion equations \cite{wheatcraft2008fractional, metzler2000random}; and characterizing non-Debye dielectric relaxation using the Prabhakar (three-parameter ML) function \cite{mainardi2015complete}. Additionally, ML functions provide analytic solutions to fractional kinetic and reaction–diffusion equations relevant in astrophysics, space sciences, and time-series analysis \cite{jose2009generalized, saxena2004generalized}. 

From this broader perspective, replacing the Boltzmann factor with its ML generalization provides a systematic way of embedding long-memory and non-equilibrium features directly into statistical weights of microstates. This approach is closely analogous to Tsallis statistics, which introduces non-extensive entropy to model fractal phase-space occupation, and to superstatistics, which accounts for fluctuations of intensive parameters. In the same spirit, the ML distribution incorporates fractional dynamics into statistical mechanics, offering a natural framework for describing anomalous transport in complex media, quantum gases with effective long-range interactions, and systems with disordered or noisy environments. Preliminary investigations in these directions are already underway, and it is anticipated that the statistical physics community—motivated by the success of Tsallis and Kaniadakis frameworks—will identify concrete applications where ML-based distributions yield experimentally testable predictions.

\section{ Mittag-Leffler Function As a Generalized Distribution}\label{3}



\begin{figure}[t]
	\includegraphics[scale=0.5]{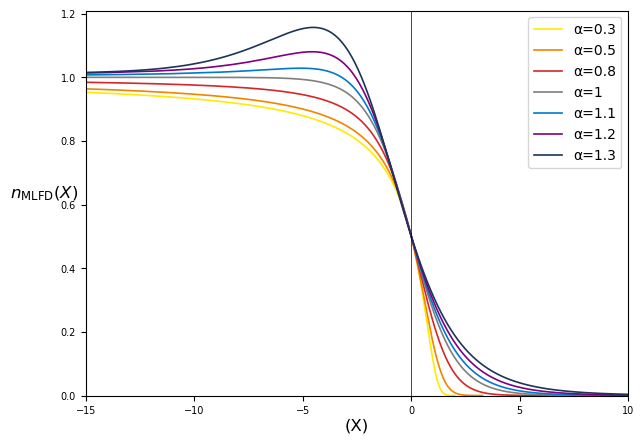}
	\captionsetup{justification=raggedright, singlelinecheck=false}
	\caption{%
		The occupation number $n_{\text{ML}}(x)$ associated with the MLFD distribution is presented as a function of the variable $x$ for several values of the parameter $\alpha$. The solid orange curve corresponds to the standard FD distribution, obtained when $\alpha = 1$.
	}
	\label{nMLFD}
\end{figure}

\begin{figure}[t]
	\includegraphics[scale=0.5]{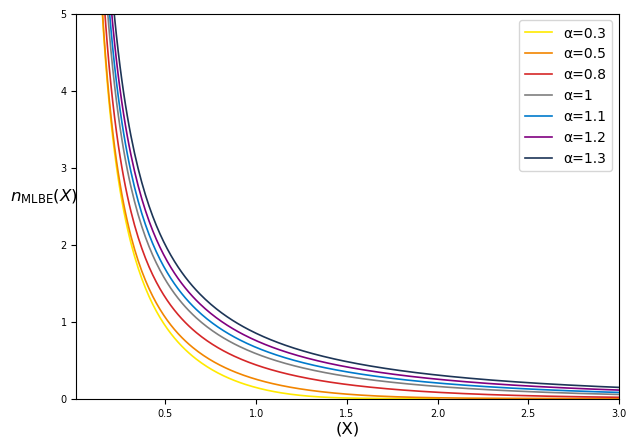}
	\captionsetup{justification=raggedright, singlelinecheck=false}
	\caption{%
		The occupation number $n_{\text{ML}}(x)$ associated with the MLBE distribution is presented as a function of the variable $x$ for several values of the parameter $\alpha$. The solid orange curve corresponds to the standard BE distribution, obtained when $\alpha = 1$.
	}
	\label{nMLBE}
\end{figure}

In equilibrium statistical mechanics, the ML, BE, and FD distributions serve as the foundational tools for describing the statistical behavior of classical and quantum particles. Each of these distributions is derived from the Boltzmann factor $e^{-\beta \epsilon}$, which assigns the statistical weight to microstates based on their energy. Although these distributions accurately model ideal systems in thermal equilibrium, they are often inadequate to capture phenomena associated with strong correlations, long-range interactions, or non-equilibrium dynamics. A systematic approach to extending the applicability of statistical mechanics involves generalizing the Boltzmann factor itself. In this work, we explore such a generalization using the ML function.

To explore this generalization more rigorously, we begin by introducing the ML function in detail. This  special function, which arises in fractional calculus and complex analysis, generalizes the exponential function and is defined via the power series
\begin{equation}
	E_{\alpha}(X) = \sum_{k=0}^{\infty} \frac{X^k}{\Gamma[\alpha k + 1]},
\end{equation}
	where $\Gamma$ is Gamma function and  \( \alpha > 0 \). In the limit \( \alpha = 1 \), the function reduces to the standard exponential \( E_1(X) = e^X \). Moreover, when \( \alpha \gg 1 \), \( E_{\alpha}(X) \) asymptotically approaches 1, reflecting a uniform limiting behavior. 
	The one-parameter form \( E_\alpha(X) \) seamlessly interpolates between exponential and power-law behaviors. Further generalizations to multi-parameter versions, such as \( E_{\alpha,\beta}(X) \) and \( E_{\alpha,\beta}^\gamma(X) \), enhance this flexibility, enabling more accurate modeling of memory effects and anomalous diffusion in complex systems~\cite{mittag1903nouvelle, haubold2011mittag}.

Having outlined its mathematical foundation, we now recall the application of the ML function in classical statistical mechanics, where it has been used to generalize the MB distribution ~\cite{seifi2025intrinsic}. We introduced the MLMB distribution by generalizing the exponential function in the classical MB distribution to the ML function. This modification enabled a smooth interpolation between classic and quantum (fermion-like and boson-like) behaviors, effectively capturing interaction-induced deviations from classical statistics. In particular, the MLMB distribution exhibits a transition between the classical behavior and the fermion-like (or boson-like) behavior depending on the value of the parameter $\alpha$. When $0 < \alpha < 1$, the system displays fermionic traits, including negative thermodynamic curvature indicative of repulsive interactions similar to the Pauli exclusion principle. For $1 < \alpha \leq 1.5$, it shows bosonic characteristics, with attractive interactions signaling positive curvature and the emergence of Bose-Einstein condensation as the fugacity approaches a critical value. This work demonstrated the versatility of the ML function in extending classical statistical mechanics beyond standard assumptions.

Building upon this classical extension, we next consider how the ML function can be incorporated into well-known quantum statistical mechanics. Specifically, we generalize the standard BE and FD distributions by replacing the exponential term with the ML function, \( E_\alpha(X) \), giving the MLBE and MLFD distributions, respectively. These are defined as
\begin{equation}\label{2}
	n_\alpha(X) = \frac{1}{E_\alpha(X) + a},
\end{equation}

	where the parameter $a$ takes the value $-1$ for MLBE, $+1$ for MLFD and  $+1$ for MLBE and $0$. Here, $X$ is defined as $\beta(\epsilon - \mu) = x - \ln z$, with $\mu$ representing the chemical potential, $x = \beta \epsilon$, and $z = e^{\beta \mu}$ denoting the fugacity. This formulation incorporates the parameter $\alpha$ directly into the quantum statistical distributions and, within appropriate limits, reproduces the BE, FD, and MB cases.

ML-based statistical distributions provide a flexible framework for describing effective interactions in quantum systems via a deformation of standard statistical mechanics. Within the formalism of thermodynamic geometry, the sign and magnitude of the thermodynamic curvature $R$ quantify intrinsic statistical correlations.

In the MLMB classical distribution~\cite{seifi2025intrinsic}, the deformation parameter $\alpha$ governs these effective interactions:
\begin{itemize}
	
\item \textbf{$\alpha = 1$}: The distribution reduces to the classical MB case. The thermodynamic curvature vanishes, consistent with the behavior of a noninteracting ideal gas.
\item \textbf{$0 < \alpha < 1$}: The curvature becomes negative, indicating effective repulsive interactions. This behavior is analogous to fermionic systems, where the Pauli exclusion principle enforces a statistical repulsion.
\item \textbf{$\alpha > 1$}: The curvature is positive, signaling effective attractive interactions, reminiscent of bosonic systems approaching BE condensation.
\end{itemize}
Importantly, the parameter $\alpha$ in the ML framework does not introduce explicit interactions. Instead, it modifies the underlying statistics. This deformation manifests in the thermodynamics as if interactions were present. Thus, the MLBE and MLFD distributions Eq. (\ref{2}) serve as a powerful phenomenological tool: they capture the net thermodynamic consequences of complex microscopic dynamics without requiring explicit modeling of those interactions.

This approach parallels other generalized statistical frameworks. For instance, the Tsallis entropic parameter $q$ and the Kaniadakis deformation parameter $\kappa$~\cite{adli2019nonperturbative,mehri2020thermodynamic}similarly tune effective interaction strength and the sign of thermodynamic curvature. The MLMB distribution extends these ideas by enabling a continuous transition between fermion-like (repulsive) and boson-like (attractive) regimes through the single parameter $\alpha$.

In summary, one may employ Eq. (\ref{2}) to analyze an interacting system by treating the deformation parameter $\alpha$ as a fitting parameter. Suppose that an experimental system exhibits thermodynamic properties, such as curvature signatures, anomalous heat capacity, or shifted phase-transition temperatures, consistent with some $\alpha \neq 1$. In that case, its behavior is effectively equivalent to that of an ideal gas obeying ML statistics with that parameter. This provides a simplified, yet comprehensive,e,e, phenomenological description of complex interacting systems.

As shown in Fig. ~\ref{nMLFD}, the MLFD distribution yields an occupation number that remains positive for all values of \( \alpha \), approaching unity as \( X \to -\infty \) and vanishing as \( X \to +\infty \). It is noteworthy that for the MLFD distribution with $\alpha > 1$, the distribution exhibits anomalous behavior: in the negative domain of $X$, it temporarily exceeds unity before asymptotically approaching 1.

In contrast, Fig.~\ref{nMLBE} shows that while the MLBE distribution also satisfies \( n(X) \to 0 \) as \( X \to +\infty \), it exhibits divergence near \( X = 0 \) and produces negative values for \( X < 0 \). These negative values of distribution are unphysical, as the occupation number must remain non-negative to ensure physical consistency. The MLFD distribution satisfies this criterion for all \( X \) and the entire range of \( \alpha \). However, the MLBE distribution imposes a constraint on the fugacity, requiring it to remain within the range $0 \leq z \leq 1$ to exclude unphysical results. At the critical value $1 < z$, the function $n(X)$ diverges, reflecting the characteristic behavior of the standard BE distribution, where BE condensation occurs~\cite{griffin1996bose,zare2012condensation,mirza2011condensation}.

\section{Thermodynamic Quantities of MLBE and MLFD Distributions}\label{4}

In the framework of statistical mechanics, the internal energy \( U \) and the total particle number \( N \) are fundamental thermodynamic quantities to characterize non-interacting quantum gases. Consider a system confined in a \( D \)-dimensional box of volume \( L^D \), where particles are distributed according to an occupation function \( n(\epsilon) \), with \(\epsilon\) representing the energy. The energy-momentum relationship is governed by the dispersion relation \( \epsilon = a p^\sigma \), where \( a \) is a constant, \( p \) denotes the momentum, and \( \sigma \) specifies the scaling exponent between energy and momentum. For non-relativistic particles\( \sigma = 2 \), reducing the relation to the classical kinetic energy expression \( \epsilon = p^2 / 2m \) with \( a = 1/2m \).
Under these assumptions, the internal energy and total particle number can be expressed as integrals over energy, weighted by the occupation function and the density of states at that energy, \(\Omega(\epsilon)\):
\begin{equation}
	\begin{split}
		U &= \int_{0}^{\infty} \epsilon \, n(\epsilon) \, \Omega(\epsilon) \, d\epsilon, \\
		N &= \int_{0}^{\infty} n(\epsilon) \, \Omega(\epsilon) \, d\epsilon,
	\end{split}
	\label{uN_mb}
\end{equation}
where \(\Omega(\epsilon)\) is the density of available quantum states as a function of energy \(\epsilon\).
The density of states generally takes the form
\begin{equation}
	\Omega(\epsilon) = A^D \, \epsilon^{\frac{D}{\sigma} - 1},
	\label{omega_general}
\end{equation}
where \( A \) is a system-dependent constant. In the case of a three-dimensional non-relativistic system, setting \( A = 1 \) reduces the density of states to $\Omega(\epsilon) = \epsilon^{1/2},$ which describes the density of states as a function of energy \(\epsilon\).

In the context of generalized quantum statistics governed by the MLBE or MLFD distributions, the thermodynamic quantities \( U \) and \( N \) are expressed through integrals involving the ML function \( E_\alpha(X) \) as follows:
\begin{equation}
	\begin{split}
		& U = \beta^{-\frac{5}{2}} \int_0^\infty \frac{x^{3/2}}{E_\alpha(X) + a} \, dx, \\
		& N = \beta^{-\frac{3}{2}} \int_0^\infty \frac{x^{1/2}}{E_\alpha(X) + a} \, dx,
	\end{split}
	\label{uN_ml_gen}
\end{equation}

For compactness and analytical utility, we define the generalized integral function:
\begin{equation}
	\int_0^\infty \frac{x^n}{E_\alpha(X) + a} \, dx = \mathcal{F}^a_{n, \alpha}(z).
\end{equation}
This function facilitates analytical developments and numerical computations.

Thus, the internal energy and total particle number can be succinctly written as
\begin{equation}
	\begin{split}
		& U = \beta^{-5/2} \mathcal{F}^a_{3/2,\alpha}(z), \\
		& N = \beta^{-3/2} \mathcal{F}^a_{1/2,\alpha}(z).
	\end{split}
	\label{uN_compact_gen}
\end{equation}

This formulation provides a unified and efficient framework for analyzing quantum gases under generalized statistics, the MLBE and MLFD.

\section{Thermodynamic Curvature of the MLBE and MLFD statistics}\label{5}

Thermodynamic geometry, as developed notably by Ruppeiner and Weinhold, provides a differential geometric structure to the space of thermodynamic parameters \cite{ruppeiner1979thermodynamics, weinhold1975metric}. Within this framework, the space of equilibrium states is treated as a Riemannian manifold, where thermodynamic fluctuations and interactions can be examined through the properties of a metric tensor. The Ruppeiner metric is derived by taking the negative second derivative of entropy with respect to extensive thermodynamic variables, such as internal energy, volume, and particle number. Alternatively, in the energy representation proposed by Weinhold, the metric is obtained from the Hessian of the internal energy with respect to its natural variables.

Applying Legendre transformations to fundamental thermodynamic potentials enables us to construct the equivalent metric structures based on various potentials, such as the Helmholtz and Gibbs free energies. Furthermore, the Fisher information metric—another significant geometric tool—can be introduced through the second derivatives of the logarithm of the partition function with respect to the intensive parameters \cite{ruppeiner1995riemannian, janyszek1990riemannian}:
\begin{equation}
	g_{ij} = \frac{\partial^2 \ln \mathcal{Z}}{\partial \beta^i \partial \beta^j},
	\label{fi}
\end{equation}
where the parameters for a two dimensional parameter space are denoted by \( \beta^1 = \beta \) and \( \beta^2 = \gamma =-\mu/k_B T\), and \( \mathcal{Z} \) is the grand canonical partition function.

In systems characterized by two thermodynamic degrees of freedom, the associated geometry is two-dimensional. Janyszek and Mrugała demonstrated that when a metric is derived directly from a thermodynamic potential, the scalar curvature of the parameters space can be expressed in terms of the potential’s second and third derivatives \cite{janyszek1989riemannian}. In two dimensions, the Ricci scalar curvature is computed via:
\begin{equation}
	R = \frac{
		\begin{vmatrix}
			g_{\beta \beta} & g_{\gamma \gamma} & g_{\beta \gamma} \\
			g_{\beta \beta,\beta} & g_{\gamma \gamma,\beta} & g_{\beta \gamma,\beta} \\
			g_{\beta \beta,\gamma} & g_{\gamma \gamma,\gamma} & g_{\beta \gamma,\gamma}
		\end{vmatrix}
	}{
		2 \left( \begin{vmatrix}
			g_{\beta \beta} & g_{\beta \gamma} \\
			g_{\beta \gamma} & g_{\gamma \gamma}
		\end{vmatrix} \right)^2
	},
	\label{rrr}
\end{equation}
where, $g_{ij,k}$ is the derivative of the elements of metric tensor with respect to the thermodynamics parameters.

For an ideal classical gas described by the MB statistics, the thermodynamic curvature vanishes. In contrast, quantum ideal gases exhibit non-zero curvature: it is positive for BE statistics and negative for FD statistics. These signs reflect effective statistical interactions—attractive in the case of bosons and repulsive for fermions \cite{ruppeiner1995riemannian,janyszek1990riemannian,oshima1999riemann}. The divergence of curvature near a critical fugacity (e.g., \( z \to 1 \) for BE) can signal the presence of phase transitions such as Bose-Einstein condensation.  The theoretical framework of geometric thermodynamics has been addressed in earlier studies \cite{mohammadzadeh2017thermodynamic,mohammadzadeh2016perturbative,mirza2011condensation,mehri2020thermodynamic,esmaili2024thermodynamic}, which provide comprehensive reviews of its geometric structure and thermodynamic implications.

In the following analysis, we extend this framework to explore the geometric structure associated with the generalized MLBE and MLFD distributions. By applying Eq. (\ref{fi}) and using the expression for the partition function relevant to the MLMB distribution, the metric components are calculated as:
\begin{equation}
	\begin{split}
		g_{\beta \beta} &= \frac{\partial^2 \ln \mathcal{Z}}{\partial \beta^2} = -\left( \frac{\partial U}{\partial \beta} \right)_{\gamma} = \frac{5}{2} \beta^{-7/2} \mathcal{F}^a_{3/2,\alpha}(z), \\
		g_{\beta \gamma} &= g_{\gamma \beta} = \frac{\partial^2 \ln \mathcal{Z}}{\partial \beta \partial \gamma} = -\left( \frac{\partial N}{\partial \beta} \right)_{\gamma} = \beta^{-5/2} \partial_{z} \mathcal{F}^a_{1/2,\alpha}(z), \\
		g_{\gamma \gamma} &= \frac{\partial^2 \ln \mathcal{Z}}{\partial \gamma^2} = -\left( \frac{\partial N}{\partial \gamma} \right)_{\beta} = z \beta^{-3/2} \partial_z \mathcal{F}^a_{1/2,\alpha}(z).
	\end{split}
	\label{g mlmb1}
\end{equation}

The derivatives required for computing the Ricci scalar are then given by:
\begin{equation}
	\begin{aligned}
		g_{\beta \beta,\beta} &= \frac{\partial}{\partial \beta} g_{\beta \beta}= -\frac{35}{4} \beta^{-\frac{9}{2}} \mathcal{F}_{\frac{3}{2},\alpha}(z), \\
		\\
		g_{\beta \beta,\gamma} &= g_{\beta \gamma,\beta} = g_{\gamma \beta,\beta} = \frac{\partial}{\partial \gamma} g_{\beta \beta} = -\frac{5}{2} \beta^{-\frac{7}{2}} \partial_{z} \mathcal{F}_{\frac{3}{2},\alpha}(z), \\
		\\
		g_{\beta \gamma,\gamma} &= g_{\gamma \beta,\gamma} = g_{\gamma \gamma,\beta} = \frac{\partial}{\partial \beta} g_{\gamma \gamma}= -\frac{3}{2}  \beta^{-\frac{5}{2}}  \partial_{z} \mathcal{F}_{\frac{1}{2},\alpha}(z), \\
		\\
		g_{\gamma \gamma,\gamma} &= \frac{\partial}{\partial \gamma} g_{\gamma \gamma}=-z \beta^{-\frac{3}{2}} (  \partial_{z} \mathcal{F}_{\frac{1}{2},\alpha}(z)  + z  \partial^{2}_{z} \mathcal{F}_{\frac{1}{2},\alpha}(z)).
	\end{aligned}
	\label{g mlmb2}
\end{equation}
By inserting the metric components from Eqs.~(\ref{g mlmb1}) and their derivatives from Eq.~(\ref{g mlmb2}) into the curvature relation given in Eq.~(\ref{rrr}), we compute the scalar curvature associated with the thermodynamic geometry of the MLBE and MLFD statistics. The resulting behavior of the curvature is illustrated in the following figures, elucidating the geometric effects induced by the underlying generalized statistical frameworks. These plots depict the thermodynamic curvature \(R\) as a function of the fugacity \(z\) for various values of the ML parameter \(\alpha\), clearly demonstrating its crucial role in determining the system’s thermodynamic behavior.  

\begin{figure}[t]
	\includegraphics[scale=0.5]{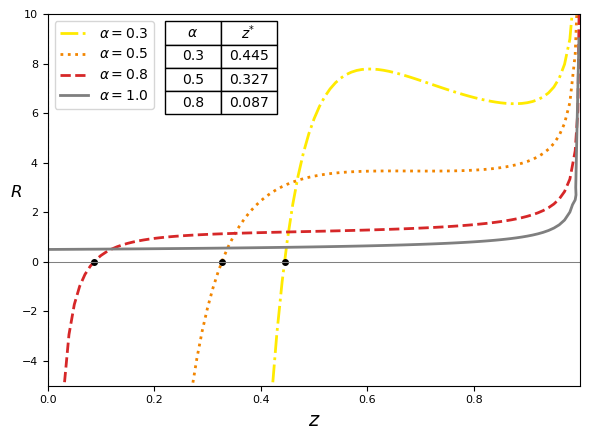}
	
	\captionsetup{justification=raggedright, singlelinecheck=false} 
	\caption{%
		The thermodynamic curvature associated with the MLBE distribution is presented as a function of fugacity for the interval $0 < \alpha \leq 1$, under isothermal conditions ($\beta = 1$). Dashed lines correspond to specific cases where $\alpha = 0.3, 0.5, 0.8$, and the respective values of $z^{*}$ are indicated by solid circles for each value of $\alpha$.
	}
	\label{RMLBEkam}
\end{figure}

In the MLBE distribution, for $\alpha < 1$, the system displays a fugacity-dependent crossover between fermionic and bosonic characteristics. At a specific point \(z = z^*_{\alpha}\), indicated by a solid dot in Fig.~\ref{RMLBEkam}, the thermodynamic curvature changes sign. For fugacity values less than \(z^*_{\alpha}\), the system displays fermionic characteristics, marked by effective repulsive interactions. In contrast, for \(z > z^*_{\alpha}\), the system transitions into the bosonic regime, characterized by attractive effective interactions between particles. This crossover, along with the dependence of \(z^*_{\alpha}\) on the parameter \(\alpha\), is illustrated in Fig.~\ref{RMLBEkam}.

\begin{figure}[t]
	\includegraphics[scale=0.5]{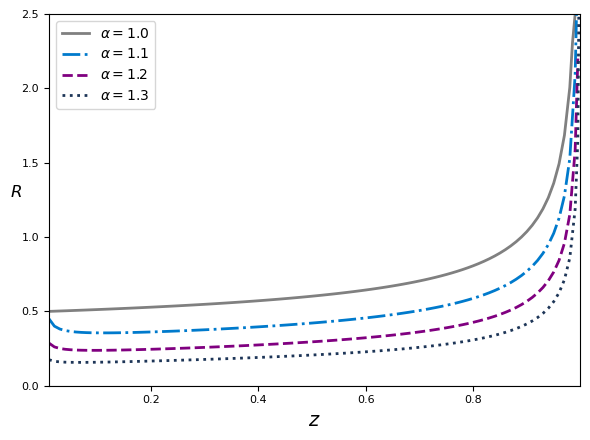}
	
	\captionsetup{justification=raggedright, singlelinecheck=false} 
	\caption{%
		Thermodynamic curvature of an MLBE distribution as a function of fugacity, plotted for the range \((1 < \alpha)\) under isothermal conditions (\(\beta = 1\)). 
		Dashed lines represent the values \(\alpha = 1.1, 1.2, 1.3\).
	}
	\label{RMLBEbish}
\end{figure}

In contrast, when \(\alpha > 1\), the MLBE distribution exhibits predominantly bosonic behavior for all values of \(z\), as indicated by the consistently positive thermodynamic curvature shown in Fig.~\ref{RMLBEbish}. Moreover, irrespective of the value of \(\alpha\), the MLBE distribution features a critical point at \(z^{\alpha}_c =z^{\alpha}_c =1\), where the thermodynamic curvature diverges. This divergence signals the onset of a phase transition analogous to Bose-Einstein condensation, as discussed in Refs.~\cite{janyszek1990riemannian,mirza2011condensation}. The critical fugacity \(z_c\) thus marks the emergence of macroscopic quantum phenomena associated with condensation.

\begin{figure}[t]
	\includegraphics[scale=0.5]{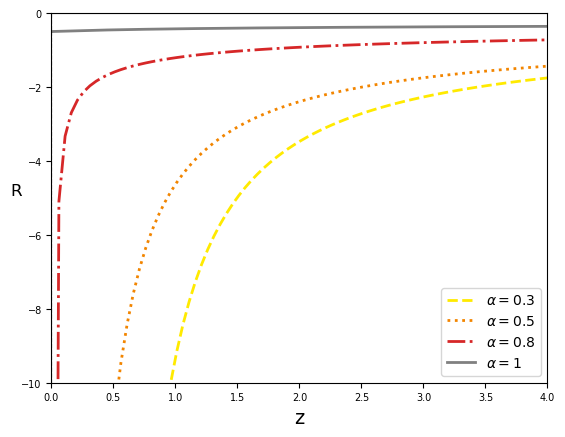}
	
	\captionsetup{justification=raggedright, singlelinecheck=false} 
	\caption{%
		Thermodynamic curvature of an MLFD distribution as a function of fugacity, plotted for the range \(1 < \alpha\) under isothermal conditions (\(\beta = 1\)). 
		Dashed lines represent the values \(\alpha = 0.3, 0.5, 0.8\).
	}
	\label{RMLFDkam}
\end{figure}

\begin{figure}[t]
	\includegraphics[scale=0.5]{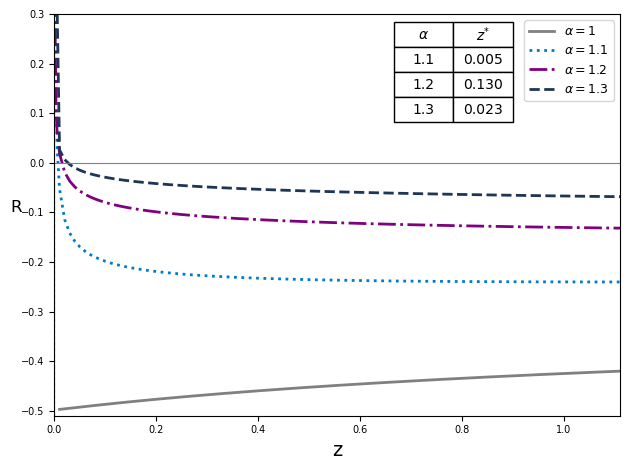}
	
	\captionsetup{justification=raggedright, singlelinecheck=false} 
	\caption{%
		Thermodynamic curvature of an MLFD distribution as a function of fugacity, plotted for the range \((1 < \alpha)\) under isothermal conditions (\(\beta = 1\)). 
		Dashed lines represent the values \(\alpha = 1.1, 1.2, 1.3\).
	}
	\label{RMLFDbish}
\end{figure}

Turning to the MLFD distribution, a similar classification emerges based on \(\alpha\). For \(\alpha < 1\), the system exhibits purely fermionic behavior: the thermodynamic curvature remains negative across all values of \(z\), starting from strongly negative values at small \(z\) and becoming less negative as \(z\) increases, see Fig. (\ref{RMLFDkam}). In contrast, for $\alpha > 1$, the thermodynamic curvature $R$ initially assumes positive values at very small values of $z$. At a specific threshold $z = z^*_{\alpha}$, as summarized in the accompanying table in Fig.~\ref{RMLFDbish}, the thermodynamic curvature undergoes a sign change. Beyond this point, $R$ rapidly transitions to negative values and gradually stabilizes with a mildly decreasing slope as $z$ increases, as shown in Fig.~\ref{RMLFDbish}.

\section{Heat Capacity of MLBE and MLFD Statistics} \label{6}

As discussed previously, in the generalized MLBE statistics, the fugacity must satisfy the condition \( z \leq z_c = 1 \) to ensure the physical validity of the distribution, that is, to maintain non-negative occupation numbers. At \( z_c  \), the system exhibits a divergence in thermodynamic curvature, signaling the onset of a phase transition. This critical behavior is associated with singularities in fundamental thermodynamic quantities. In this section, we explore the implications of this divergence on thermodynamic response functions, with a particular focus on the heat capacity at constant volume, \( C_v \). This quantity is defined as the temperature derivative of the internal energy at fixed volume:

\begin{equation}\label{Cv_mb}
	\frac{C_v}{N k_B} = \left( \frac{\partial U}{\partial T} \right)_V.
\end{equation}

Using the expressions for the internal energy and particle number derived from MLBE statistics, Eq. (\ref{uN_compact_gen}), we obtain a generalized expression for the heat capacity in terms of the integral functions $\mathcal{F}^a_{n,\alpha}(z)$

\begin{equation} \label{cv mlbefd}
	\frac{C_v}{N k_B} = \frac{5 \, \mathcal{F}^a_{3/2,\alpha}(z)}{2 \, \mathcal{F}^a_{1/2,\alpha}(z)} - \frac{9 \, \mathcal{F}^a_{1/2,\alpha}(z)}{2 \, \mathcal{F}^a_{1/2,\alpha}(z)}.
\end{equation}

The normalized heat capacity in Eq.~(\ref{cv mlbefd}) applies to both MLBE and MLFD statistics and incorporates the influence of the parameter \( \alpha \). These generalized quantum frameworks exhibit a temperature-dependent heat capacity due to the implicit dependence of the fugacity  on temperature.

To investigate this dependence more explicitly, we analyze the variation of \( C_v / N k_B \) as a function of the scaled temperature \( T / T_c^\alpha \), where \( T_c^\alpha \) denotes the critical temperature associated with each value of \( \alpha \).

Determining the critical temperature \( T_c^\alpha \) for a given \( \alpha \), corresponding to the critical fugacity \( z_c \), is essential for characterizing the phase structure of the system. This temperature depends explicitly on \( \alpha \) and must be evaluated for a fixed particle number \( N \). Using Eq.~(\ref{uN_compact_gen}), it is given by:
\begin{equation} \label{Tc_alpha}
	T_{c}^{\alpha} = \frac{h^{2}}{2\pi m k_{B}} \left( \frac{N}{V \, \mathcal{F}^{a}_{1/2,\alpha}(z_c^\alpha)} \right)^{2/3}.
\end{equation}

\begin{figure}[t]
	\includegraphics[scale=0.5]{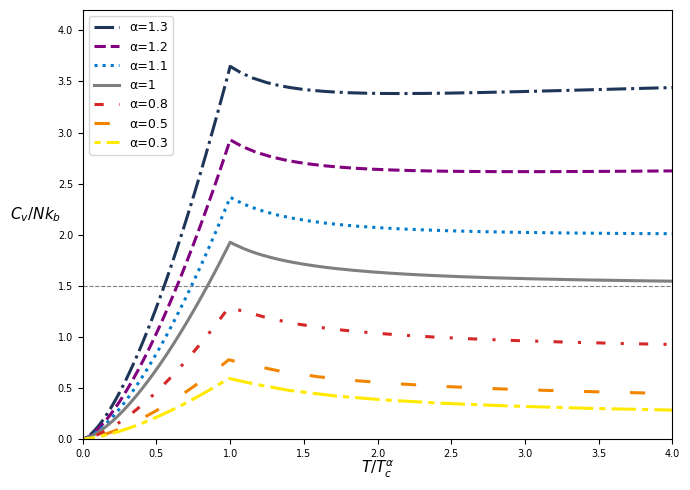}
	
	\captionsetup{justification=raggedright, singlelinecheck=false} 
	\caption{%
		The heat capacity of MLBE at fixed volume as a function of temperature, plotted for the range. Dashed lines represent several values of the parameter $\alpha$.
	}
	\label{CVMLBE}
\end{figure}

\begin{figure}[t]
	\includegraphics[scale=0.5]{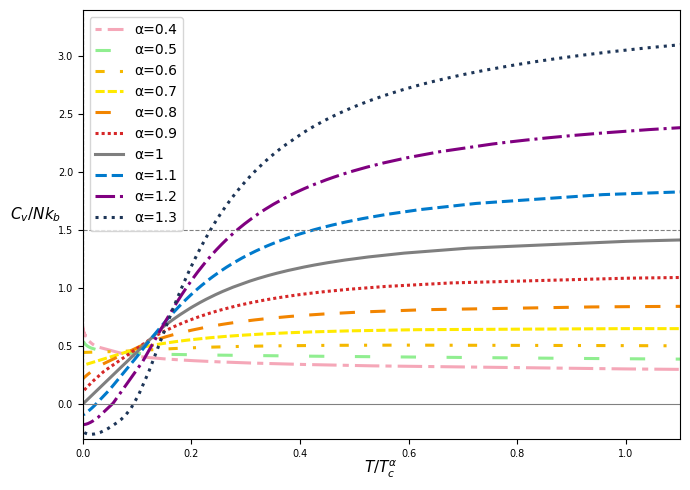}
	\captionsetup{justification=raggedright, singlelinecheck=false} 
	\caption{%
		The heat capacity of MLFD at fixed volume as a function of temperature, plotted for the range. Dashed lines represent several values of the parameter $\alpha$.
	}
	\label{CVMLFD}
\end{figure}

The temperature dependence of the normalized heat capacity \( C_v / N k_B \) for the MLBE statistics, obtained from Eq.~(\ref{cv mlbefd}), is presented in Fig.~\ref{CVMLBE}. The resulting curve reveals a phase transition at the temperature \( T/T^{\alpha}_c = 1 \), characterized by a non-analytic behavior, akin to the Bose-Einstein condensation observed in the conventional BE framework. 

However, unlike the standard BE case, where the heat capacity approaches the classical limit of \( 3/2 \) at high temperatures, the MLBE system exhibits qualitatively distinct behavior. In the high-temperature limit, \( C_v \) depends explicitly on the value of the ML parameter \(\alpha\). Even in this regime—where classical and quantum statistics typically converge—the MLBE distribution retains signatures of its generalized nature, emphasizing the influence of \(\alpha\) on the thermodynamic behavior.

Notably, the high-temperature value of the heat capacity is a monotonically increasing function of \(\alpha\). In particular, for \(\alpha > 1\), the high-temperature heat capacity exceeds the classical (and  quantum statistics) limit, whereas for \(\alpha < 1\), it remains below that limit.

In the framework of MLFD statistics, the temperature dependence of the normalized heat capacity, \( C_v / N k_B \), is depicted in Fig. \ref{CVMLFD}. For values of \(\alpha < 1\), the heat capacity curve undergoes a notable qualitative transition, characterized by a reversal in curvature and a shift from a decreasing to an increasing trend. This transition occurs around \(\alpha \approx 0.6\), which serves as an approximate threshold that marks the point of behavioral change. Specifically, for \(\alpha \lesssim 0.6\), the heat capacity decreases with increasing temperature, whereas in the range \(\alpha \approx 0.6\) to 1, it exhibits a gradual increase. In all cases with \(\alpha < 1\), the heat capacity begins from a positive value at low temperatures.

At \(\alpha = 1\), corresponding to the classical Fermi--Dirac statistics, the heat capacity starts from zero and, with increasing temperature, asymptotically approaches the classical limit of \(3/2\). This behavior is consistent with well-established results from classical statistical mechanics.

For $\alpha > 1$, the MLFD statistics reveals a pronounced thermodynamic anomaly: the heat capacity becomes negative at low temperatures, indicating thermodynamic unstable and deviation from conventional statistical behavior. As the temperature increases, the heat capacity rapidly transitions to positive values, followed by a more gradual rate of increase. Notably, unlike in conventional models, the heat capacity does not converge to a fixed constant in the high-temperature limit for \(\alpha > 1\). This non-standard behavior highlights the unique statistical features inherent to the MLFD distribution when \(\alpha\) exceeds unity.

\subsection{Negative Heat Capacity in the MLFD Distribution} 

In classical thermodynamics, the heat capacity is typically positive, reflecting the general expectation that an increase in the internal energy of a system leads to a corresponding increase in temperature. This behavior is consistent with extensive systems in the thermodynamic limit and is well-described within the canonical ensemble. However, in certain finite or nonextensive systems—particularly those influenced by long-range interactions or constrained by limited degrees of freedom—this assumption may break down.

Negative heat capacity arises when the entropy function \( S(U) \), defined in terms of the total energy \( U \), exhibits a locally convex region; that is, when \( \frac{d^2 S}{dU^2} > 0 \). Given that the microcanonical temperature is defined by \( \frac{1}{T} = \frac{dS}{dU} \), this condition implies that an increase in energy can lead to a decrease in temperature, thereby violating the canonical ensemble's stability criteria. Such behavior is commonly associated with first-order phase transitions in finite systems, where the coexistence of distinct phases leads to non-monotonic features in the caloric curve.

In the present study, we demonstrated that negative heat capacity can emerge directly from a generalization of standard quantum statistical distributions, without requiring specific interaction types. By extending the conventional distribution functions, the negative heat capacity observed here arises as a purely statistical consequence. This suggests that the phenomenon is not necessarily restricted to specialized physical mechanisms but may instead be broadly accessible within a wider class of systems governed by generalized statistics.

Given the statistical origin of negative heat capacity revealed by the MLFD distribution, it is natural to inquire whether similar thermodynamic behavior has been observed experimentally. Although the results presented in this study are derived from a theoretical framework, analogous features have been reported across a variety of experimental contexts—particularly in finite, isolated, or nonextensive systems. In the following section, we review selected experimental studies in which negative heat capacity has been documented. These include investigations of nuclear multifragmentation, atomic and molecular clusters, and self-gravitating astrophysical systems. Collectively, these examples underscore the broader thermodynamic significance of negative heat capacity and provide insight into the physical conditions under which this phenomenon may arise.

\subsection{Evidence for Negative Heat Capacity in Finite Systems}

Empirical evidence for negative heat capacity was first reported by D’Agostino \textit{et al.}~\cite{d1999thermodynamical}, who studied multifragmentation in Au + Au collisions at 35~A.MeV. Using event-by-event analysis within the microcanonical ensemble, they reconstructed the excitation energy and temperature, revealing a distinct region of negative heat capacity around 5~A.MeV. This observation aligns with theoretical predictions that finite systems undergoing a first-order phase transition may develop a convex entropy branch and exhibit negative microcanonical heat capacity.

Further support was provided by D’Agostino \textit{et al.}~\cite{d2000negative} through a detailed analysis of caloric curves and energy fluctuations during the liquid-gas phase transition in finite nuclear systems. Their results demonstrated a clear thermodynamic signature in the microcanonical caloric curve, highlighting the necessity of microcanonical analysis for capturing such noncanonical behavior.

Evidence from cluster physics was presented by Schmidt \textit{et al.}~\cite{schmidt2001negative}, who investigated Na\(_{147}^{+}\) atomic clusters near their melting transition. Using photofragmentation mass spectrometry and reconstruction of microcanonical caloric curves, they observed a “backbending” in the temperature-energy relation—an unmistakable sign of negative heat capacity. This was associated with a convex entropy profile satisfying \( \frac{d^2S}{dU^2} > 0 \), confirming the presence of negative heat capacity in isolated mesoscopic systems.

On the theoretical side, Padmanabhan~\cite{padmanabhan1990statistical} showed that self-gravitating systems with long-range interactions inherently exhibit negative heat capacity within the microcanonical ensemble. Employing toy models and mean-field approximations, the study revealed that these systems transition between kinetic-energy-dominated and potential-energy-dominated phases through an intermediate regime characterized by negative heat capacity. This regime corresponds to virial equilibrium and disappears in the canonical ensemble, where it is replaced by a discontinuous phase transition. These findings emphasize the inequivalence of statistical ensembles and the central role of the microcanonical framework in describing systems with long-range interactions.

Additional theoretical support comes from the microcanonical formalism developed by Gross~\cite{gross2001microcanonical}, which investigates phase transitions in small systems beyond the thermodynamic limit. It is argued that the microcanonical heat capacity can be extracted from fluctuations in the kinetic energy of nuclear fragments. In multifragmenting nuclei, large fluctuations in kinetic energies, combined with a convex entropy curvature, lead to negative microcanonical heat capacity values.

Inspired by D’Agostino \textit{et al.}’s approach, this method relates the observed kinetic energy fluctuations to the curvature of the entropy surface, \( \frac{d^2S}{dU^2} \), which is not directly measurable experimentally. In model systems such as the two-dimensional Hamiltonian Mean Field model, this curvature can be explicitly calculated and shown to produce negative heat capacity in the transition region. Negative total heat capacity emerges from a positive curvature in the logarithm of the accessible phase space volume, corresponding to a convex intruder in the entropy surface—a well-known signature of first-order phase transitions in finite systems.

Together, these experimental and theoretical results provide robust, multidisciplinary evidence for the existence of negative heat capacity. The phenomenon is now recognized as a generic feature of finite, isolated systems and serves as a diagnostic of underlying nonextensive or first-order phase transition behavior. Moreover, these findings highlight the limitations of canonical ensemble treatments for small or nonextensive systems and underscore the essential role of microcanonical thermodynamics in revealing subtle features, such as negative heat capacity, which are otherwise suppressed in ensemble-averaged descriptions.

	\section{Comparative Analysis with Tsallis and Kaniadakis Statistics}\label{7}

To contextualize the present results, it is useful to compare the thermodynamic geometry and heat capacity of the ML-based statistics with those of Tsallis and Kaniadakis statistics. Each of these frameworks generalizes the exponential function through a parameter, recovering standard statistical behavior in the appropriate limit.

\subsection{	Explanation of the Tsallis Function}

	The Tsallis function ~\cite{Tsallis1988possible,adli2019nonperturbative}, or \( q \)-entropy, is a generalization of the Boltzmann-Gibbs (BG) entropy introduced by Tsallis to describe systems with nonextensive properties. It is defined as:
	
	\[
	S_q \equiv -k_B \frac{1 - \sum_{i=1}^W p_i^q}{1 - q}, \quad q \in \mathbb{R},
	\]
	
	where \( p_i \) is the probability of the system being in microstate \( i \), \( k_B \) is the Boltzmann constant, and \( W \) is the total number of microstates. The parameter \( q \) quantifies the degree of nonextensivity. When \( q \to 1 \), the Tsallis entropy reduces to the BG entropy:
	
	\[
	S_{BG} \equiv -k_B \sum_{i=1}^W p_i \ln p_i.
	\]
	
	The Tsallis entropy is particularly useful for systems with long-range interactions, fractal structures, or other complexities that make the BG entropy inadequate. The associated \( q \)-generalized distribution function for particle statistics is:
	
	\[
	n_q(\epsilon) = \frac{1}{\left(1 + (q-1)\beta(\epsilon - \mu)\right)^{\frac{1}{q-1}} - \alpha},
	\]
	
	where \( \alpha = 1 \) for Tsallis-FD, \( \alpha = -1 \) for Tsallis-BE, and \( \alpha = 0 \) for classic Tsallis-MB statistic. This distribution reduces to the standard FD, BE, and MB distributions when \( q \to 1 \).

	\subsection{	 Explanation of the Kaniadakis Function}
	The Kaniadakis function~\cite{kaniadakis2001non,mehri2020thermodynamic}, or $\kappa-$exponential, is a one-parameter deformation of the standard exponential function, introduced by Kaniadakis to generalize BG statistics. It is defined as:
	
	\[
	\exp_{\kappa}(x) = \left(\sqrt{1 + \kappa^2 x^2} + \kappa x\right)^{1/\kappa},
	\]
	
	where \(\kappa\) is the deformation parameter. In the limit \(\kappa \to 0\), the $\kappa-$exponential reduces to the ordinary exponential function, recovering standard statistics. The $\kappa-$exponential is used to define generalized distribution functions for particles obeying $\kappa-$statistics, such as the $\kappa-$MB, $\kappa-$BE, and $\kappa-$FD distributions. These distributions modify the statistical behavior of particles, introducing non-extensive effects that can describe systems with long-range interactions or non-equilibrium states.

		\subsection{	 	Thermodynamic Geometry}

In the Tsallis framework ~\cite{adli2019nonperturbative}, the thermodynamic curvature exhibits behavior largely consistent with the standard statistics. For Tsallis-BE with $q > 1$, the curvature is positive, indicating attractive interactions, and diverges at fugacity $z = 1$, a signature of BE condensation. Increasing $q$ reduces the magnitude of $R$, suggesting enhanced stability. In contrast, Tsallis-BE gases display negative curvature, consistent with repulsive interactions, while nonextensive Tsallis-BE gases remain interaction-free, with $R = 0$. Thus, the Tsallis parameter modifies the intensity of statistical interactions but does not alter their qualitative nature.
	
	By comparison, the Kaniadakis framework ~\cite{mehri2020thermodynamic} reveals more nuanced geometric behavior. In $\kappa-$deformed MB systems, $R$ deviates from the classical zero value and becomes negative, indicating $\kappa-$induced repulsion. $\kappa-$deformed Fermi systems consistently exhibit negative curvature, as expected from fermionic repulsion. For $\kappa-$deformed Bose systems, however, the curvature is more intricate: below a critical fugacity $z^*$, $R$ is negative, resembling fermionic interactions, while above $z^*$, it becomes positive, reflecting bosonic attraction. At $z = 1$, the divergence of $R$ signals BE condensation.
	
	The ML statistics further generalize this picture, allowing not only the modification of interaction strengths but also qualitative crossovers between bosonic- and fermionic-like behavior. In the MLBE case, when $\alpha < 1$, the thermodynamic curvature changes sign at a fugacity $z = z_\alpha^*$: the system exhibits fermionic-like repulsion ($R < 0$) for $z < z_\alpha^*$ and bosonic-like attraction ($R > 0$) for $z > z_\alpha^*$. For $\alpha > 1$, however, the system is purely bosonic, with $R > 0$ for all fugacity values. In all cases, a divergence at $z_c^\alpha = 1$ marks the onset of BE condensation.
	
	Similarly, in the MLFD case, the curvature remains negative for all $z$ when $\alpha < 1$, reflecting persistent fermionic repulsion. For $\alpha > 1$, however, the curvature is positive at small fugacity, transitions through a threshold $z_\alpha^*$, and becomes negative at larger $z$, indicating a crossover from bosonic-like to fermionic-like effective behavior. Once again, divergence at $z_c^\alpha = 1$ signals condensation.

	A comparative examination of Tsallis, Kaniadakis, and ML statistics highlights both commonalities and essential differences in their thermodynamic geometries. The Tsallis framework represents a conservative extension: although the deformation parameter \$q\$ modifies the magnitude of thermodynamic curvature and thereby the strength of effective interactions, the fundamental distinction between bosonic attraction and fermionic repulsion remains unchanged. Kaniadakis statistics, in contrast, introduce more intricate behavior. The $\kappa$-deformation not only adjusts interaction strengths but also generates novel features, such as repulsion in Maxwell–Boltzmann systems and a dual character in Bose systems, where the curvature changes sign at a critical fugacity. ML statistics extend this generalization even further: by allowing genuine sign reversals of the thermodynamic curvature as functions of fugacity and the parameter $\alpha$, they permit authentic crossovers between bosonic- and fermionic-like regimes, offering the broadest and most flexible framework of the three.
	
The central point is that, as demonstrated in Reference ~\cite{seifi2025intrinsic}, the MLMB statistic —a classical ML-based statistic— exhibits distinct interaction behaviors that are contingent upon the value of the parameter $\alpha$. Specifically, for $\alpha > 1$, the interactions are attractive, whereas for $\alpha < 1$, they are repulsive. When $\alpha = 1$, the results recover the classical MB statistics, corresponding to a non-interacting system. These findings suggest that MLMB statistics inherently encompass both attractive and repulsive interactions, in contrast to the classical Tsallis statistic, which operates under a purely non-interacting framework, and the classical Kaniadakis statistic, which presumes solely repulsive thermodynamic curvature.

		\subsection{Heat Capacity}

The heat capacity offers a complementary perspective on the thermodynamic behavior of systems, particularly in the vicinity of phase transitions. Within the framework of Tsallis-BE statistics, the heat capacity at constant volume, $C_V$, exhibits non-analytic behavior at the critical temperature $T_c^{(q)}$. Below this transition, $C_V \propto T^{3/2}$, whereas above $T_c^{(q)}$, the system adheres to the modified thermodynamic relations characteristic of Tsallis statistics. The non-analyticity diminishes with increasing $q$, becoming effectively smooth in the limit of large $q$.

A different behavior emerges in the context of ML-based statistics. In the MLBE case, the high-temperature limit of $C_V$ explicitly depends on the deformation parameter $\alpha$. For $\alpha > 1$, $C_V$ exceeds the classical value of $3/2$, while for $\alpha < 1$, it remains below this benchmark.

The MLFD statistic demonstrates even more unconventional features. For $\alpha > 1$, the heat capacity becomes negative at low temperatures -indicating thermodynamic instability- before transitioning to positive values at higher temperatures.


In the context of classical MLMB statistics, as demonstrated in~\cite{seifi2025intrinsic}, the heat capacity reduces to the standard classical value, $C_V = 3/2$, when $\alpha = 1$. For $\alpha > 1$, the system exhibits behavior analogous to the MLBE case with $\alpha > 1$, while for $\alpha < 1$, the ratio $C_V / (N k_B)$ diverges at low temperatures and asymptotically approaches zero at higher temperatures, without ever reaching it exactly.

    \section{The MLBE statistics Debye solid}\label{8}

	Following the approach presented in~\cite{kittel2018introduction}, we adopt the Debye framework. In the present work, we extend this formulation by replacing the conventional BE statistics with the generalized MLBE statistics.
    
	In the Debye approximation, the sound velocity is assumed to be constant for each polarization type, akin to a classical elastic continuum. The dispersion relation is given by
	\begin{equation}
		\omega = v k ,
	\end{equation}
	where $v$ represents the constant sound velocity..
	
	The corresponding density of states is expressed as
	\begin{equation}
		D(\omega) = \frac{V \omega^2}{2 \pi^2 v^3} .
	\end{equation}
	
In a specimen comprising \( N \) primitive cells, the total number of acoustic phonon modes is equal to \( N \). The cutoff frequency, denoted as \( \omega_D \), is defined by the following equation:
	
	\begin{equation}
		\omega_D^3 = \frac{6 \pi^2 v^3 N}{V}.
	\end{equation}
	
	The corresponding cutoff wavevector in $k$-space is
	\begin{equation}
		k_D = \frac{\omega_D}{v} = \left( \frac{6 \pi^2 N}{V} \right)^{1/3}.
	\end{equation}
	
	The Debye model excludes modes with $k > k_D$, so that the total number of allowed modes exactly matches the degrees of freedom of a monatomic lattice.
	
	The total phonon energy is generalized as
	\begin{equation}
		U = \int_0^{\omega_D} D(\omega)\, n_\alpha(\omega)\, \hbar \omega \, d\omega ,
	\end{equation}
	where $n_\alpha(\omega)$ is the Mittag--Leffler occupation function
	\begin{equation}
		n_\alpha(\omega) = \frac{1}{E_\alpha(\hbar \omega / k_B T) - 1}.
	\end{equation}
	
	Substituting the density of states gives
	\begin{equation}
		U^\alpha =(\frac{V \omega^2}{2 \pi^2 v^3})  \int_0^{\omega_D} \frac{\hbar \omega}{E_\alpha(\hbar \omega / k_B T) - 1} \, d\omega.
	\end{equation}
	
	Assuming phonon velocity is independent of polarization, we multiply by a factor of $3$:
	\begin{equation}\label{uDebye}
		U^\alpha  = \frac{3 V \hbar}{2 \pi^2 v^3} \int_0^{\omega_D} \frac{\omega^3}{E_\alpha(\hbar \omega / k_B T) - 1} \, d\omega.
	\end{equation}
	
By introducing the dimensionless variable
	\begin{equation}\label{xd}
		x = \frac{\hbar \omega}{k_B T}, \quad x_D = \frac{\theta}{T},
	\end{equation}
	The internal energy can be rewritten as
	\begin{equation}
		U^\alpha  = \frac{3 V k_B^4 T^4}{2 \pi^2 v^3 \hbar^3} \int_0^{x_D} \frac{x^3}{E_\alpha(x) - 1}\, dx.
	\end{equation}
	
	For a system with $N$ atoms, this yields
	\begin{equation}
		U^\alpha  = 9 N k_B T \left( \frac{T}{\theta} \right)^3 \int_0^{x_D} \frac{x^3}{E_\alpha(x) - 1} \, dx ,
	\end{equation}
	where the Debye temperature is

	\begin{equation}\label{deb_mlbe}
\theta=\frac{\hbar v }{ k_B}.(\frac{6\pi^2 N}{V})^{1/3}.
\end{equation}

	The heat capacity at constant volume is determined by differentiating the internal energy, as expressed in Eq. \ref{uDebye}, with respect to temperature, in conjunction with Eq. \ref{xd} placement, leading to the following expression:
\begin{equation}\label{deb_mlbe}
	C_V^\alpha  = 9 N k_B \left( \frac{T}{\theta} \right)^3 
	\int_0^{x_D} \frac{x^4 E_\alpha(x)}{\left(E_\alpha(x) - 1 \right)^2} \, dx.
\end{equation}

\begin{figure}[t]
 	\includegraphics[scale=0.5]{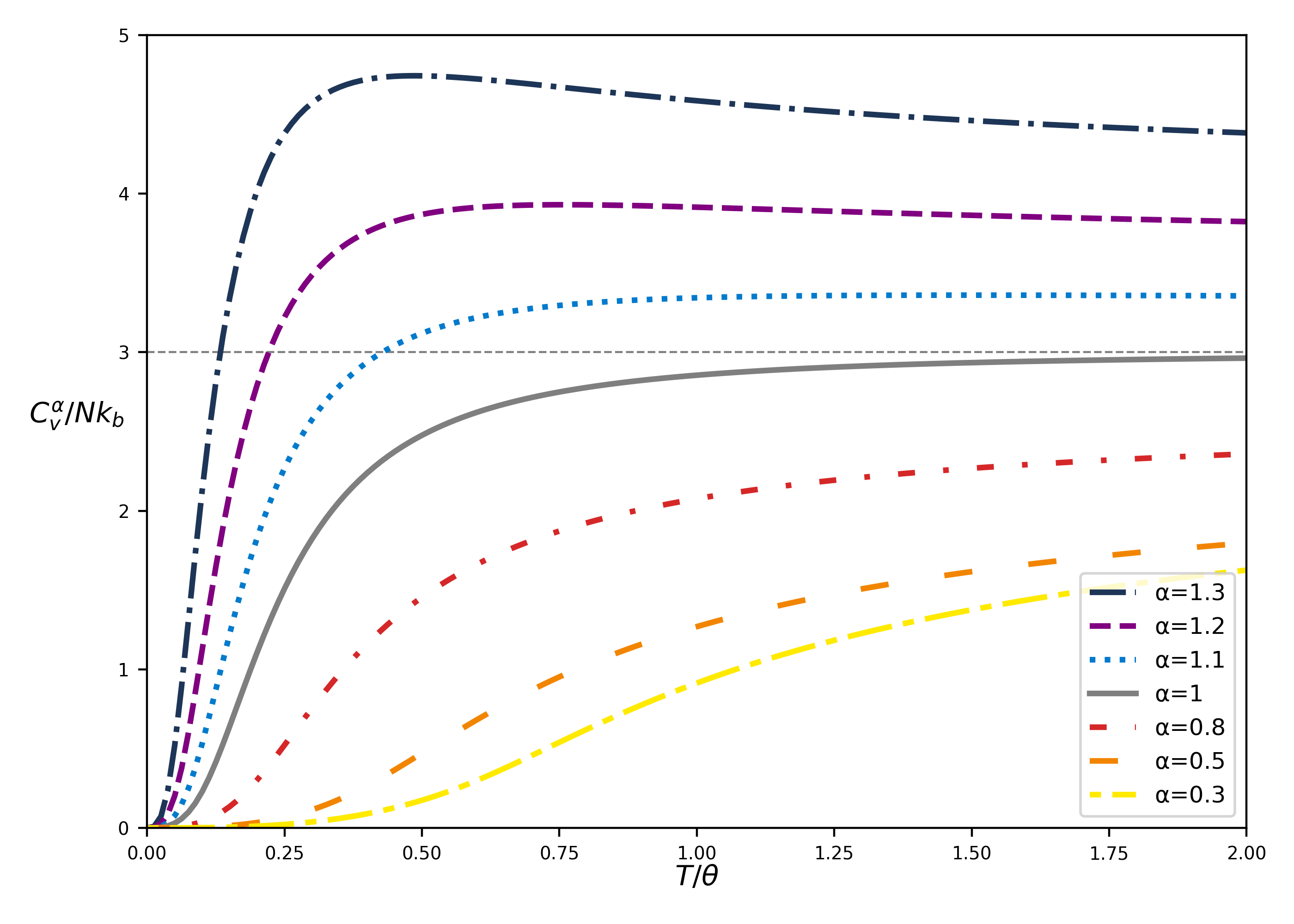}
 	
 	\captionsetup{justification=raggedright, singlelinecheck=false} 
 	\caption{%
 	Heat capacity at constant volume of the MLBE with $(\mu=0)$ as a function of $T/\theta$, based on the Debye approximation. The curves correspond to different values of the parameter $\alpha$, with $\theta$ fixed at 400.
}
 	\label{CVBEDebye}
 \end{figure}

The curves in the Fig. ~\ref{CVBEDebye} diagram show that the heat capacity begins at zero and increases monotonically with temperature. At low temperatures, this increase is relatively steep, but the growth rate diminishes at higher temperatures until the heat capacity approaches an asymptotic constant value. The case of $\alpha = 1$ corresponds to the standard case. For $\alpha < 1$, the heat capacity remains consistently below the standard Debye model, yielding curves positioned beneath the $\alpha = 1$ across the entire temperature range. In contrast, for $\alpha > 1$, the heat capacity exceeds the standard case, resulting in curves that lie above the case of $\alpha = 1$. This classification highlights the role of the parameter $\alpha$ in shifting the magnitude of the heat capacity relative to the standard Debye model.
 
\begin{figure}[t]
	\includegraphics[scale=0.5]{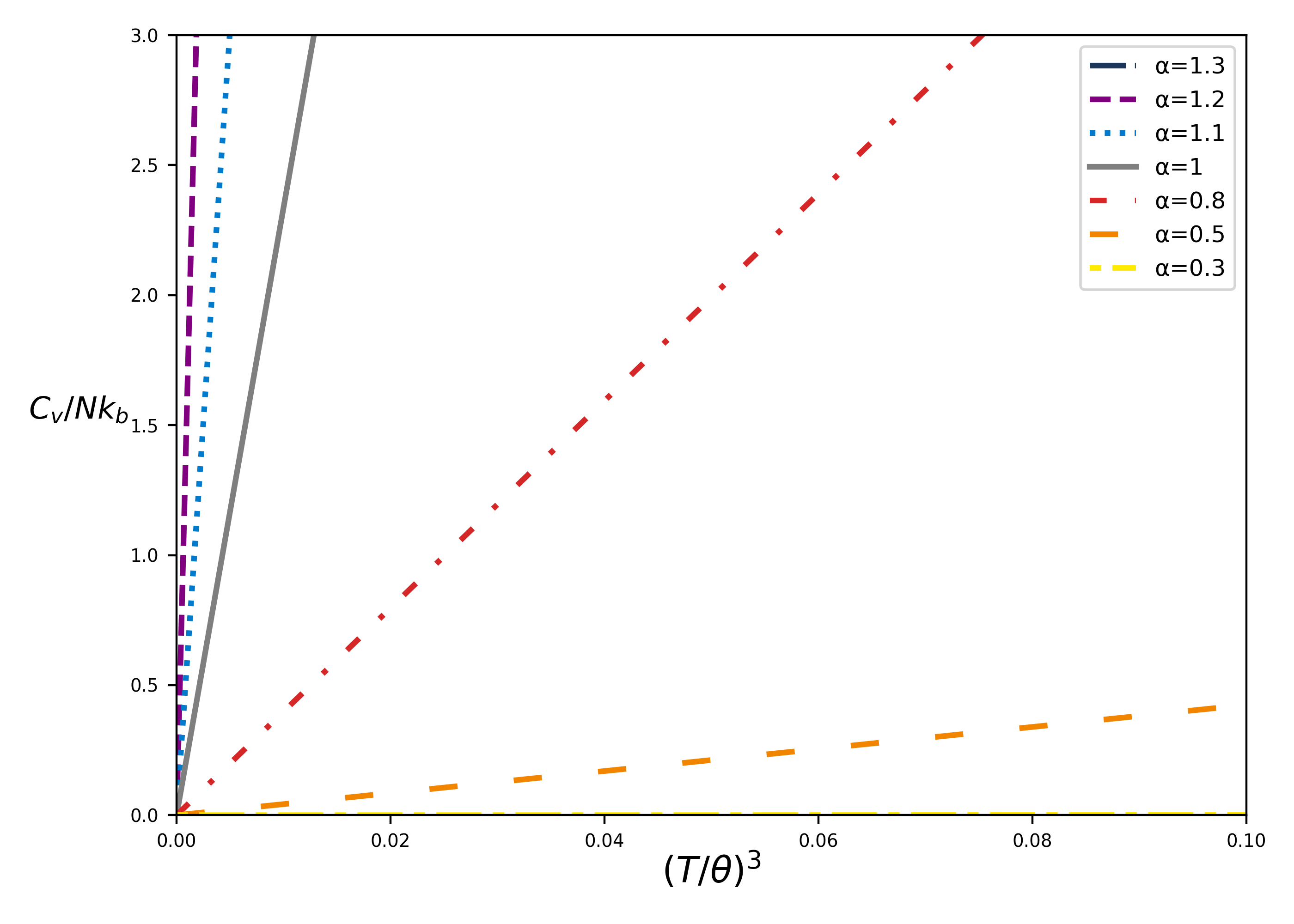}
	\captionsetup{justification=raggedright, singlelinecheck=false} 
	\caption{%
	Heat capacity at constant volume of the MLBE with $(\mu=0)$ as a function of $(T/\theta)^3$ in the low-temperature regime, based on the Debye approximation. The curves represent varying values of the parameter $\alpha$, with $\theta$ fixed at 400. 
 	}
	\label{LOWERML1}
\end{figure}

\begin{itemize}
	\item \textbf{High-temperature regime ($x_D \ll 1$, i.e., $T \gg \theta$):} 
In this regime, the argument of the Mittag--Leffler function is sufficiently small to allow the linear approximation
\begin{equation}
	E_\alpha(x) \approx 1 + \frac{x}{\Gamma(1+\alpha)}.
\end{equation}

Consequently, the integral in \eqref{deb_mlbe} can be approximated as

	\begin{equation}\label{deb2}
		\int_0^{x_D} \frac{x^4 E_\alpha(x)}{(E_\alpha(x)-1)^2} \, dx 
		\approx \int_0^{x_D} x^2 \left[\Gamma(1+\alpha)\right]^2 \, dx .
	\end{equation}
	Substituting this expression into \eqref{deb_mlbe} leads to
	\begin{equation}
		C_V^\alpha  \approx 3 N k_B \left[\Gamma(1+\alpha)\right]^2.
	\end{equation}
 Therefore, in the classical high-temperature regime for $\alpha = 1$, the heat capacity is consistent with the Dulong--Petit law:
	\begin{equation}
		C_V^1  \longrightarrow 3 N k_B.
	\end{equation}

Table~\ref{tab:results} presents the corresponding values of $C_V$ in the high-temperature regime for different values of $\alpha$. These values are in agreement with the trends observed in Figure~\ref{CVBEDebye}, which is generated from equation \eqref{deb_mlbe}.

\begin{table}[t]
	\centering
	\setlength{\arrayrulewidth}{0.5pt}   
	\renewcommand{\arraystretch}{1.5}     
	\setlength{\tabcolsep}{12pt}          
	\begin{tabular}{|c|c|}
		\hline
		\textbf{$\alpha$} & \textbf{$C_V^\alpha /N k_B $} \\
		\hline
		0.3 & 2.4164 \\
		\hline
		0.5 & 2.3562 \\
		\hline
		0.8 & 2.6024 \\
		\hline
		1 & 3.0000 \\
		\hline
		1.1 & 3.2854 \\
		\hline
		1.2 & 3.6419 \\
		\hline
		1.3 & 4.0837 \\
		\hline
	\end{tabular}
	\caption{Heat capacity at constant volume for the MLBE with $(\mu=0)$ in the high-temperature regime, computed using the Debye approximation for various values of $\alpha$, with $\theta$ fixed at 400.}
	\label{tab:results}
\end{table}

	\item \textbf{Low-temperature regime ($x_D \gg 1$, i.e., $T \ll \theta$):} 
	In this limit, the integral in \eqref{deb_mlbe} can be extended to infinity:
	\begin{equation}\label{deb3}
		\int_0^{\infty} \frac{x^4 E_\alpha(x)}{(E_\alpha(x)-1)^2} \, dx = D_\alpha.
	\end{equation}
	Inserting this into \eqref{deb_mlbe} yields
	\begin{equation}
		C_V^\alpha  \approx 9 N k_B \left( \frac{T}{\theta} \right)^3 D_\alpha.
	\end{equation}

	This $T^3$ dependence of the heat capacity is consistent with the prediction of the standard Debye bosonic model. The agreement confirms that the generalized Mittag--Leffler formulation reproduces the expected low-temperature scaling of the conventional Debye theory.
	
	Figure~\ref{LOWERML1} depicts the variation of constant-volume heat capacity within the Debye model across different values of the deformation parameter $\alpha$. The heat capacity demonstrates a linear relationship with the normalized temperature $(T/\theta)^3$, with an increasing slope of the curves corresponding to higher values of $\alpha$.
\end{itemize}

\subsection{Generalized Debye Model with $q$-Deformed}

	The thermodynamic anomalies identified in the MLBE and MLFD frameworks—namely, the anomalous high-temperature heat capacity in the MLBE case and the emergence of negative heat capacity in the MLFD scenario—are not phenomena unique to ML-based statistics. Analogous features have been documented within other generalized statistical formalisms, particularly those grounded in $q$-deformed quantum statistics derived from the Tamm–Dancoff (TD) oscillator algebra. In recent work, Chung and Algin ~\cite{chung2024q} examined the thermodynamic properties of a $q$-deformed Debye model, wherein phonons are treated as excitations obeying a TD-type $q$-deformed bosonic algebra. 
	
Chung and Algin derived the number distribution for the present $q$-phonon gas model as
	
	\begin{equation}\label{81}
		n_{i,q}=\frac{q}{e^{\beta E_i}-q},
	\end{equation}
	
They obtain the heat capacity for the q-deformed Debye solid as follows  \cite{chung2024q}
	
\begin{equation}\label{85}
	C_V^q = 9 N k_B \left( \frac{T}{\theta} \right)^3 \int_0^{x_D} \frac{ e_q(x) \, x^4 }{ \left( e_q(x) - 1 \right)^2 }  dx,
	\end{equation}
	
In their study, they further analyzed the behavior of the Debye model in both the low- and high-temperature regimes.

In the following, we present a comparison of the results derived from the $q$-deformed Debye model with those obtained from the MLBE framework.
	
\subsection{Comparative Analysis}

The preceding sections established the heat capacity within the Debye framework generalized using the MLBE statistics. In parallel, $q$-deformed Debye models~\cite{chung2024q} provide an alternative generalized-statistics formulation in which phonons obey a TD-type $q$-deformed bosonic algebra. 

A direct comparison of the two approaches reveals both similarities and distinctions:

\begin{itemize}
	\item \textbf{High-temperature regime:} 
	In the MLBE framework, the high-temperature heat capacity is approximated as
	\begin{equation}
		C_V^\alpha \approx 3 N k_B [\Gamma(1+\alpha)]^2\quad \text{(MLBE)},
	\end{equation}
	which for $\alpha = 1$ recovers the classical Dulong--Petit value $C_V \rightarrow 3 N k_B$. In contrast, the $q$-deformed Debye solid exhibits \cite{chung2024q}
	\begin{equation}
		C_V^q \approx 36\, N k_B \, \ln\!\left(\frac{q-1}{q}\right)  \quad \text{($q$-deformed)},
	\end{equation}
	which deviates from the classical limit and may become negative for $q>1$, indicating anomalous thermodynamic behavior not accessible in the conventional Debye model.
    
    Thus, while both generalized frameworks modify the classical heat capacity, the MLBE approach preserves the Dulong--Petit consistency for $\alpha = 1$, whereas the $q$-deformed model introduces potentially nonphysical high-temperature anomalies.

	\item \textbf{Low-temperature regime:} 
	At low temperatures, both frameworks recover the expected $T^3$ scaling of the heat capacity:
	\begin{equation}
			C_V^\alpha \approx 9 N k_B \left(\frac{T}{\theta}\right)^3 D_\alpha \quad \text{(MLBE)},
	\end{equation}
	\begin{equation}
		c_V^q \approx 216\, N k_B \left(\frac{T}{\theta}\right)^3 \mathrm{Li}_4(q) \quad \text{($q$-deformed)}.
	\end{equation}
		where the polylogarithm function $\mathrm{Li}_4(q)$ is defined by the series $	\mathrm{Li}_s(z) = \sum_{k=1}^{\infty} z^k/k^s $ \cite{chung2024q}.

		This agreement indicates that both generalized statistical models recover the traditional Debye scaling in the low-temperature regime, thereby conforming to the established bosonic phonon model. Both models maintain the characteristic $T^3$ scaling at lower temperatures.	

\end{itemize}
These characteristics demonstrate that deviations from classical thermodynamic behavior are not exclusive to any single generalized-statistics approach; rather, they represent a fundamental aspect of generalized distributions.

\section{Conclusion and Outlook}\label{9}

In this work, we have introduced and explored two novel generalizations of quantum statistical distributions - the MLBE and MLFD distributions - by generalizing the exponential function in the standard BE and FD formulations to the ML function. This substitution introduces a parameter\( \alpha \), which offers a tunable mechanism to capture non-equilibrium systems.

We derived the thermodynamic properties of MLBE and MLFD systems, including expressions for internal energy and particle number, and analyzed their behavior in the thermodynamic limit. Using thermodynamic geometry, we examined the curvature of the associated state spaces. We uncovered rich geometric structures, including curvature singularities signaling phase transitions and crossover behavior between fermionic and bosonic characteristics depending on the value of \( \alpha \).

The analysis reveals that the MLBE distribution possesses a critical fugacity $z_c = 1$, at which the thermodynamic curvature diverges, signaling the presence of a phase transition analogous to Bose-Einstein condensation. In the regime $\alpha < 1$, a crossover behavior is observed, wherein the system transitions from effective fermionic to bosonic characteristics. Conversely, the MLFD distribution exhibits a consistently negative thermodynamic curvature for $\alpha > 1$, reflecting the presence of effective repulsive interactions typically associated with fermionic systems.

To further investigate the thermodynamic consequences of the generalized distributions, we analyzed the heat capacity as a function of temperature. In both the MLBE and MLFD frameworks, the heat capacity exhibits a temperature dependence that is strongly influenced by the parameter $\alpha$, revealing nontrivial deviations from conventional statistical behavior.
Most notably, in the MLFD statistics case \( \alpha > 1 \), the system exhibits a pronounced thermodynamic anomaly: the heat capacity becomes negative at low temperatures, signaling thermodynamic instability and a significant deviation from standard FD behavior. Intriguingly, this anomaly is mirrored in the behavior of the MLFD distribution function itself: for \( \alpha > 1 \), it temporarily exceeds unity in the negative domain of the variable \( X \), before asymptotically approaching one. The coincidence of these anomalies in both the statistical distribution and the heat capacity underscores a fundamental deviation from standard statistical behavior in this regime.

The findings establish ML statistics as an important formalism, positioning it alongside well-established frameworks such as those of Tsallis and Kaniadakis. When considered within the broader context of statistical theories, the ML approach demonstrates notable versatility. In contrast to Tsallis statistics, which primarily alter the intensity of inherent statistical interactions, and Kaniadakis statistics, which can induce sign changes in curvature for bosons, the ML formalism enables genuine transitions between effective bosonic and fermionic behavior. This ability to qualitatively modify the nature of statistical interactions through the adjustment of the parameter $\alpha$ makes the ML distributions a more adaptable and comprehensive framework for modeling complex systems.

The application of the Debye model using MLBE phonon statistics further substantiates the physical validity of this approach. It successfully reproduces the canonical $T^3$ law at low temperatures and predicts $\alpha$-dependent corrections to the high-temperature Dulong-Petit limit.

In conclusion, the ML generalization of quantum statistics offers a robust and unified formalism that effectively captures a wide range of thermodynamic phenomena. The parameter $\alpha$ acts as a tuning mechanism for introducing specific non-equilibrium or interaction effects, providing a valuable tool for theorists modeling complex quantum systems. Promising avenues for future research include: exploring the implications of these distributions in relativistic quantum fields and astrophysical contexts; investigating the use of multi-parameter ML functions for enhanced descriptive control; applying the MLBE framework to study anomalous phonon transport in disordered solids; and seeking experimental evidence of ML statistics in ultracold quantum gases or nanoscale systems where tunable disorder and non-Markovian dynamics are prevalent.

\section*{Acknowledgments}

The authors gratefully acknowledge Professor Morteza Natagh Najafi for his invaluable discussions and insights regarding the physical applications of ML functions. His guidance greatly enriched the understanding and presentation of the topics covered in this work.  
Furthermore, the authors sincerely thank the two anonymous referees whose constructive comments and suggestions significantly improved the quality and clarity of this manuscript.

	\bibliographystyle{apsrev4-2}
\bibliography{re}

\end{document}